\documentclass[12pt]{article}

\usepackage{times}
\usepackage{graphics}
\usepackage{amssymb}

\renewcommand{\thefootnote}{\fnsymbol{footnote}}

\newcommand{\nc}{\newcommand}
\newcommand{\rnc}{\renewcommand}

\rnc{\baselinestretch}{1.24}    
\rnc{\arraystretch}{1.24}       

\makeatletter
\rnc{\theequation}{\thesection.\arabic{equation}}
\@addtoreset{equation}{section}
\makeatother

\nc{\be}{\begin{equation}}
\nc{\ee}{\end{equation}}
\nc{\bea}{\begin{eqnarray}}
\nc{\eea}{\end{eqnarray}}
\nc{\xx}{\nonumber \\}

\nc{\eq}[1]{(\ref{#1})}
\nc{\newcaption}[1]{\centerline{\parbox{6in}{\caption{#1}}}}

\nc{\fig}[3]{
\begin{figure}
\centerline{\epsfxsize=#1\epsfbox{#2.eps}}
\newcaption{#3. \label{#2}}
\end{figure}
}

\nc{\np}[3]{Nucl. Phys. {\bf B#1} (#2) #3}
\nc{\pl}[3]{Phys. Lett. {\bf #1B} (#2) #3}
\nc{\prl}[3]{Phys. Rev. Lett.{\bf #1} (#2) #3}
\nc{\prd}[3]{Phys. Rev. {\bf D#1} (#2) #3}
\nc{\ap}[3]{Ann. Phys. {\bf #1} (#2) #3}
\nc{\prep}[3]{Phys. Rep. {\bf #1} (#2) #3}
\nc{\rmp}[3]{Rev. Mod. Phys. {\bf #1} (#2) #3}
\nc{\cmp}[3]{Comm. Math. Phys. {\bf #1} (#2) #3}
\nc{\mpl}[3]{Mod. Phys. Lett. {\bf #1} (#2) #3}
\nc{\cqg}[3]{Class. Quant. Grav. {\bf #1} (#2) #3}
\nc{\jhep}[3]{J. High Energy Phys. {\bf #1} (#2) #3}

\def\CN{{\cal N}}

\def\IP{\mathbb{P}}
\def\IR{\mathbb{R}}

\def\a{\alpha}

\def\g{\gamma}
\def\d{\delta}
\def\e{\epsilon}

\def\k{\kappa}
\def\l{\lambda}
\def\m{\mu}

\def\t{\tau}

\def\w{\omega}

\def\D{\Delta}

\def\L{\Lambda}
\def\S{\Sigma}

\def\half{\frac{1}{2}}

\def\goto{\rightarrow}

\def\del{\nabla}

\def\p{\partial}

\def\Tr{{\rm Tr}}

\def\va{\vec{a}}

\def\vn{\vec{n}}
\def\vx{\vec{x}}
\def\vy{\vec{y}}

\begin{document}

\begin{titlepage}
\hfill\parbox{4cm}
{hep-th/0205279}

\vspace{20mm}
\centerline{\Large \bf PP-wave/Yang-Mills Correspondence:} 
\vspace{3mm}
\centerline{\Large \bf An Explicit Check}

\vspace{10mm}
\begin{center}
Youngjai Kiem$^a$\footnote{ykiem@muon.kaist.ac.kr},
Yoonbai Kim$^b$ \footnote{yoonbai@skku.ac.kr},
Sangmin Lee$^b$\footnote{sangmin@newton.skku.ac.kr},
and Jaemo Park$^c$\footnote{jaemo@physics.postech.ac.kr}
\\[10mm]
{\sl $^a$ Department of Physics, KAIST, Taejon 305-701, Korea}
\\[2mm]
{\sl $^b$ BK21 Physics Research Division and Institute of Basic
Science}
\\
{\sl Sungkyunkwan University, Suwon 440-746, Korea}
\\[2mm]
{\sl $^c$ Department of Physics, POSTECH, Pohang 790-784, Korea}

\end{center}

\thispagestyle{empty}
\vskip 30mm

\begin{abstract}
\noindent
We present an explicit evidence that shows the correspondence 
between the type IIB supergravity in the pp-wave background 
and its dual supersymmetric Yang-Mills theory at the interaction level.  
We first construct the cubic term of the light-cone interaction 
Hamiltonian for the dilaton-axion sector of the supergravity.  
Our result agrees with the corresponding part of 
the light-cone string field theory (SFT) and 
furthermore fixes its previously undetermined $p^+$-dependent 
normalization. Adopting thus fixed light-cone SFT, we compute 
the matrix elements of light-cone Hamiltonian involving three 
chiral primary states and find an agreement with a prediction 
from the dual Yang-Mills theory.

\end{abstract}

\vspace{2cm}
\end{titlepage}

\baselineskip 7mm
\renewcommand{\thefootnote}{\arabic{footnote}}
\setcounter{footnote}{0}

\section{Introduction}

While the duality between type IIB string theory 
in a pp-wave background \cite{figu,blau2,mets,blau1} and a sector of 
${\cal N} = 4$ supersymmetric Yang-Mills theory with large R charge \cite{bmn}
is in a sense a part of the AdS$_5$/CFT$_4$ correspondence, 
its many novel features make one to approach it 
from a different angle. 
The authors of Ref. \cite{bmn} succeeded in reproducing the 
string spectrum \cite{metstey} from perturbative Yang-Mills theory, 
thereby putting the duality on a firm ground at the free theory level.
Subsequent papers made some progress on the important question 
of string interactions both on the Yang-Mills theory side 
\cite{seme, bn, gross, harvmit}  
and on the string theory side \cite{spvo}.
\footnote{See Refs. \cite{papa2}-\cite{bon} 
for related recent developments.} 

String spectrum in the pp-wave background can be obtained 
from that of the full AdS, at least in principle, 
by simply taking a limit.
However, the dictionary between AdS and CFT at the interaction level 
is seemingly lost. In the AdS/CFT correspondence, the
correlators of the CFT are recovered from string theory 
(supergravity in practice) by using boundary-to-bulk propagators 
and bulk supergravity interaction vertices.  
The pp-wave limit magnifies the neighborhood of a null geodesic 
and pushes away the original boundary, so that after taking the limit
the notion of the bulk-to-boundary propagator becomes unclear.

Without clear understanding of how holography is realized 
for the pp-wave case (see however Refs. \cite{bn,das,kirit,roza}), 
one should first specify 
what physical quantities can be computed and compared 
between the two theories. The string theory in the pp-wave 
background takes the simplest form in the light-cone gauge. 
As shown in Ref. \cite{bmn}, the light-cone Hamiltonian is related 
to the AdS variables as $H_{l.c.} = \m (\D - J)$ where 
$\D$ is the AdS energy and $J$ is a component of angular momentum 
along $S^5$. In the language of Yang-Mills defined on $S^3\times \IR$, 
$\D$ is again energy and $J$ is a $U(1)$ R charge. 
Since $J$ does not change due to perturbation, 
the interaction Hamiltonians of the two theories are expected to 
be the same once appropriate observables are identified. 
That is, if matrix elements among well-defined states can be computed 
reliably in both theories, that will provide a non-trivial check 
of the duality at the interaction level. 
Note that scattering amplitudes are not well-defined in a pp-wave 
since the strings are confined by an effective gravitational potential; 
we are solving a quantum mechanics for `particles in a box.'
Note also that the quantum mechanical approach proved useful 
already in recovering the string spectrum from Yang-Mills theory \cite{bmn}.

In Ref. \cite{spvo} the cubic term of the interaction Hamiltonian 
of light-cone string field theory in a pp-wave background was determined
up to an overall function of light-cone momenta $p^+$. 
Once one fixes the normalization, one can in principle 
compute arbitrary matrix elements among three single particle states.  
On the other hand, in the course of computing the second 
order correction to the energy spectrum in the Yang-Mills theory, 
the authors of Ref. \cite{harvmit} made a concrete proposal for 
a set of matrix elements. 

In this paper, we make an attempt to build a bridge between 
the two computations. Specifically, we fix the normalization 
of the cubic Hamiltonian of Ref. \cite{spvo} 
by comparing it with a supergravity calculation. 
We then use it to compute the matrix elements 
of three chiral primary states and find that the result agrees 
with the prediction of Ref. \cite{harvmit}. 
Strictly speaking, the supergravity analysis in the present paper 
is valid for $\m p^+ \a' \ll 1$ while the Yang-Mills analysis of 
Ref. \cite{harvmit} is valid for $\m p^+ \a' \gg 1$. 
However, our experience in AdS/CFT \cite{3pt} tells us that 
the cubic interactions involving chiral primaries are strongly protected 
by supersymmetry. 
It is plausible that the nearly chiral string states 
of Ref. \cite{bmn} show only a small deviation from their chiral cousins.
In any case, confirmation of the proposal of Ref. \cite{harvmit} 
for chiral primaries should be a first consistency check
as a prelude to a full-fledged comparison involving 
the string states of Ref. \cite{bmn}. 
Verification of the proposal for them will be reported elsewhere.

This paper is organized as follows.
In section 2, we carefully perform the light-cone quantization 
of the dilaton-axion system of IIB supergravity in a pp-wave background. 
The fact that the cubic Hamiltonian is nearly the same 
as the one in the flat spacetime enables us to determine it uniquely 
using an argument partially based on Lorentz symmetry.
In Section 3, we compare the cubic Hamiltonian $H_3$ of Ref. \cite{spvo} 
with that of section 2 and fix the normalization of the latter. 
We then compute the matrix elements of $H_3$ for three chiral primary states, 
and compare them with the proposal of Ref. \cite{harvmit}.
In an appendix, we present a computation of the same matrix elements  
in the full AdS space. We again find an agreement, but the derivation 
is not fully faithful.

\section{Light-cone quantization of IIB supergravity
in a pp-wave background}

One of the most efficient ways of constructing the 
light-cone Hamiltonian of superstring theory
is to solve the constraints from (super)symmetries, 
which uniquely fixes the Hamiltonian in the flat background 
\cite{extgra,brink}.  
In pp-wave backgrounds whose symmetry generators are inherited from
those of AdS space \cite{machi,hats}, however, the counterparts of the 
flat space generators $J^{+-}$ and $J^{-I}$ are not present. 
This absence leaves the overall $p^+$-dependent factor of the interaction 
Hamiltonian undetermined, barring us from performing the  
comparison to the corresponding proposals based on Yang-Mills theory.  
It thus appears necessary to directly construct
the light-cone Hamiltonian of the pp-wave supergravity
from the known covariant action and to compare it with the
zero mode part of the string interaction Hamiltonian.

We begin with the determination of the cubic interaction
Hamiltonian of the IIB supergravity in a pp-wave background.
The final answer that we get should be the same as the one
obtained from the
light-cone string field theory computations after fixing its overall
$p^+$-dependent normalization.  For this purpose, it is enough
to work out the simplest non-trivial case, namely, the dilaton-axion
system.  In the Einstein frame, the bosonic action of
type IIB supergravity with manifest $SL(2,R)$ invariance
is given as follows:
\begin{eqnarray}
S_{\rm IIB} & = & \frac{1}{2 \kappa^2}
\int d^{10}x \sqrt{-g} \Big( R - \frac{ \partial_{\mu} \tau
 \partial^{\mu} \bar{\tau}}{ 2 ( {\rm Im} ~ \tau )^2 }
 - \frac{{\cal M}_{ij}}{2} F_3^i \cdot F_3^j  \nonumber \\
 & & - \frac{1}{4} |\tilde{F}_5 |^2
 - \frac{\epsilon_{ij}}{4} C_4 \wedge F_3^i \wedge F_3^j \Big) ,
\label{2bsugra}
\end{eqnarray}
where
\begin{equation}
 \tau = \chi + i e^{-\phi} ~~~ , ~~~
 {\cal M}_{ij} = \frac{1}{{\rm Im} ~ \tau}
  \pmatrix{  | \tau |^2 & - {\rm Re}~ \tau \cr
           - {\rm Re}~ \tau  &  1  } ~~~ , ~~~
  F_3^i = \pmatrix{ H_3 \cr F_3 } ~ ,
\end{equation}
and
\begin{equation}
\tilde{F}_5 = F_5 - \frac{1}{2} C_2 \wedge H_3 + \frac{1}{2}
 B_2 \wedge F_3 ~ ,
\end{equation}
and the coupling constant is given by $2 \kappa^2 = (2 \pi )^7 g_s^2
\alpha^{\prime 4}$.  
As is well-known, the action given in the above is incomplete,
for the self-duality condition
\[ * \tilde{F}_5 = \tilde{F}_5 \]
should be separately imposed.  While this procedure is
easy to perform on the classical solutions, it is
difficult to implement at the quantum level.  Concentrating
on the dilaton-axion sector allows us to bypass
this subtlety.  Note that there is no coupling in
(\ref{2bsugra}) between the dilaton-axion and $\tilde{F}_5$.

Expanding (\ref{2bsugra}) up to cubic orders
for the dilaton-axion sector around $\tau = \tau' + i= i$ 
($\phi = 0$, $\chi = 0$)  
and setting other perturbations to be zero, we have
\begin{equation}
S =  \frac{1}{2 \kappa^2} \int dx^{10} \sqrt{-g}
\left( - \frac{1}{2} \nabla_{\mu} \tau \nabla^{\mu} \bar{\tau}
 - \frac{i}{2} (\tau - \bar{\tau} ) \nabla_{\mu} \tau
\nabla^{\mu} \bar{\tau}   \right) ~ , 
\end{equation}
where we suppressed the prime in $\tau'$. 
Rescaling $\tau \rightarrow
\sqrt{2} \kappa \tau$ and $\bar{\tau} \rightarrow
\sqrt{2} \kappa \bar{\tau}$ shows that $\sqrt{2} \kappa$ is 
the cubic coupling constant.
The background metric $g$ is set to that of
the pp-waves in IIB supergravity
\begin{equation}
ds^2 = g_{\mu \nu} dx^{\mu} dx^{\nu} =
-4dx^+ dx^- - \m^2 x^I x^I (dx^+)^2 + (dx^I)^2 ~ , ~~~
 \sqrt{-g} = 2 ,
\end{equation}
where $x^+$ is the light-cone `time' and $I = 1, \cdots , 8$.
Written explicitly, the action is
\begin{eqnarray}
S & = &  \int dx^+ dx^- dx^I \Big(
\frac{1}{2} \partial_+ \tau \partial_- \bar{\tau}  +
\frac{1}{2} \partial_- \tau \partial_+ \bar{\tau}
 - \frac{1}{4} \mu^2 x^I x^I \partial_- \tau \partial_-
 \bar{\tau} -   \partial_I \tau \partial_I \bar{\tau} \nonumber \\
& &
+ i \sqrt{2} \kappa (\tau - \bar{\tau} ) ( \frac{1}{2}
 \partial_+ \tau \partial_- \bar{\tau}  +
\frac{1}{2} \partial_- \tau \partial_+ \bar{\tau}
 - \frac{1}{4} \mu^2 x^I x^I \partial_- \tau \partial_-
 \bar{\tau} -  \partial_I \tau \partial_I \bar{\tau} ) \Big) ~ .
\label{sugac}
\end{eqnarray}
One might try to construct the quantizable
light-cone Hamiltonian from (\ref{sugac}) treating
$\sqrt{2} \kappa$ as a perturbation expansion
parameter. This approach, however, has a subtle problem.
The light-cone
canonical momenta are computed to be
\begin{equation}
2 \Pi_{\tau} =  \partial_- \bar{\tau} + i \sqrt{2} \kappa 
 ( \tau - \bar{\tau} )
 \partial_- \bar{\tau}  ~ ,
\end{equation}
\begin{equation}
2 \Pi_{\bar{\tau}} =  \partial_- \tau + i \sqrt{2} \kappa
( \tau - \bar{\tau} )
 \partial_- \tau ~ .
\label{pchi}
\end{equation}
The cubic interaction terms of supergravity action involve two
derivatives.  Due to them, the
canonical momentum $\Pi_{\tau}$ gets the second,
interaction-dependent term.
This behavior greatly complicates the imposition of the
standard canonical commutation relations when we try to quantize
(\ref{sugac}).  Specifically, the
cubic part of the commutation relation yields
nonlinear constraints between the $\tau$ operator and the $\bar{\tau}$
operator.
In the case of
nonabelian gauge theories
having single derivative cubic interactions, this difficulty
can be overcome as follows;
the interaction dependent part
of the canonical momentum (proportional to
$\partial_- A_I - \partial_I A_- + g_{YM} [ A_- , A_I ]~$)
vanishes upon choosing the light-cone gauge
$A_- = 0$.  In our case, a simple cure is to introduce a
nonlocal field redefinition (and its complex conjugated version)
\begin{equation}
 \tau \rightarrow \tau - \sqrt{2} \kappa
 \frac{i}{2} \frac{1}{\partial_-} \left(
 (\tau - \bar{\tau}) \partial_- \tau \right) ~, 
\label{redef}
\end{equation}
\begin{equation}
 \bar{\tau} \rightarrow \bar{\tau} - \sqrt{2}
\kappa \frac{i}{2}
  \frac{1}{\partial_-} \left(
 (\tau - \bar{\tau}) \partial_- \bar{\tau} \right)
\label{redefbar}
\end{equation}
that deletes the part of the cubic action that contributes to
$\Pi_{\tau}$.  We will then try to quantize the resulting
system.  In  (\ref{redef}), we understand
\begin{equation}
 \frac{1}{\partial_-} f(x^- ) =
 \int^{x^-} dx^{- \prime} f( x^{- \prime} )
\end{equation}
or its momentum space version
\begin{equation}
 \frac{i}{\partial_-} \rightarrow \frac{1}{2p^+} = \frac{1}{\alpha} ,
\end{equation}
where $\alpha = 2 p^+$ throughout the rest of this paper.
Strictly speaking, the field redefinition (\ref{redef}) should
be applied to nonzero modes with $p^+ \ne 0$.  One can show that
a slightly modified version of (\ref{redef}) for the zero mode $p^+ =
0$ part may be used to delete the cubic terms involving 
$\partial_+$-derivative and zero modes.

Upon performing the field
redefinition, the action (\ref{sugac}) changes to:
\begin{eqnarray}
S & = &  \int dx^+ dx^- dx^I \Big( \frac{1}{2}
\partial_+ \tau \partial_- \bar{\tau}  + \frac{1}{2}
\partial_- \tau \partial_+ \bar{\tau}
 - \frac{1}{4} \mu^2 x^I x^I \partial_- \tau \partial_-
 \bar{\tau} -  \partial_I \tau \partial_I \bar{\tau} \nonumber \\
& & + i \sqrt{2}  \kappa  \Big[ -  (\tau - \bar{\tau} )
\partial_I \tau \partial_I \bar{\tau}
+ \frac{1}{2} \partial_I \left( \frac{1}{\partial_-} 
 ( (\tau - \bar{\tau})  \partial_- \tau ) \right) \partial_I 
   \bar{\tau} \nonumber \\
 & &
+ \frac{1}{2} \partial_I \left( \frac{1}{\partial_-} ( (\tau - \bar{\tau})
   \partial_- \bar{\tau} ) \right) \partial_I \tau \Big]
 + {\rm quartic~and~higher~terms} \Big) ~ ,
\label{rsugac}
\end{eqnarray}
where we have formally used integration by parts.
The cubic interaction terms in (\ref{rsugac}) do not involve any
$\partial_+$-derivatives.  An important point is that
they do not involve any $\mu$-dependent parts either and,
in fact, they are identical to the {\em flat space} supergravity
results.  The field redefinition that removes interaction
terms having $\partial_+$-derivatives
also removes the $\mu$-dependent interaction terms (without
using integration by parts).  We propose to
take (\ref{rsugac}) as the starting point for further analysis.

The main subtlety of light-cone Hamiltonian construction is
the careful treatment of $p^+ = 0$ zero modes and the implementation
of the extra global constraint from them.  In the flat space-time case, 
Lorentz invariance emerges only after properly incorporating
the zero mode effect, as reported for example in \cite{mask}.  
What makes our problem solvable is that the zero mode
part of the action (\ref{rsugac}) is $\mu$-independent. 
The only $\mu$-dependent term in (\ref{rsugac}) involves 
$\partial_-$-derivative and vanishes for the zero modes.
Taken together, these
mean that the extra $\mu$-independent contribution to 
the light-cone Hamiltonian coming from the zero modes should
make the total Hamiltonian compatible with Lorentz invariance when $\mu = 0$.

Fortunately, for the theory at hand (\ref{rsugac}), 
the zero-mode contribution to the light-cone Hamiltonian 
which ensures Lorentz invariance can be 
inferred from an earlier work by Goroff and Schwarz \cite{gor}.
The light-cone Hamiltonian thus obtained from (\ref{rsugac}) 
is given as follows:
\begin{equation}
 H = H_2 +  H_3 ~ ,
\label{sugh}
\end{equation}
where
\begin{equation}
H_2  =   \int dx^- dx^I
\left(
  \frac{1}{4} \mu^2 x^I x^I \partial_- \tau \partial_-
 \bar{\tau} +   \partial_I \tau \partial_I \bar{\tau}
\right) ~ ,
\label{sugh2}
\end{equation}
and
\begin{eqnarray}
H_3 & = & \sqrt{2} \kappa i  \int dx^- dx^I \Big(
  \tau  \partial_I \tau \partial_I \bar{\tau}
 - \frac{1}{2} \partial_I \left( \frac{1}{\partial_-} ( \tau
   \partial_- \tau ) \right) \partial_I \bar{\tau} \nonumber \\
& &
 - \frac{1}{2}  \partial_I \left( \frac{1}{\partial_-} ( \tau
   \partial_- \bar{\tau} ) \right) \partial_I \tau
 + {\rm c.c.} \Big) + H_{GS}   ~ .
\label{sugh3}
\end{eqnarray}
The extra term $H_{GS}$ guarantees the Lorentz invariance
up to cubic terms when $\mu = 0$ and is given by 
\begin{equation}
H_{GS} = \frac{\sqrt{2} \kappa}{4i} \left(
 \tau \partial_I \tau \partial_I \bar{\tau}
- \left( \frac{\partial_I}{\partial_-} \tau \right)
 \partial_I \tau \partial_- \bar{\tau}
- \left( \frac{\partial_I}{\partial_-} \tau \right)
 \partial_I \bar{\tau} \partial_- \tau
 + \left( \frac{\partial_I \partial_I}{\partial_-^2}
\tau \right) \partial_- \tau \partial_- \bar{\tau}
\right) + {\rm c.c.} ~ .
\end{equation}
In the above equations ${\rm c.c.}$ represents the
complex conjugate.  We note that 
the relative normalization between 
$H_{GS}$ and the other terms in $H_3$ 
is uniquely fixed by requiring the eventual
Lorentz invariance when $\mu = 0$. 

It is worthwhile to give a brief review of the work by Goroff and Schwarz 
\cite{gor}. 
They considered the construction of
light-cone Hamiltonian for a nonlinear sigma model
with $SL(D-2,R)$ invariance to mimic the $D$-dimensional
`gravity' theory:
\begin{equation}
S = \int d^D x \left( K_{+-} - \frac{1}{2} \gamma^{ij}
K_{ij} + \left( \frac{\partial_j}{\partial_-} \gamma^{ij} 
\right) K_{i-} - \frac{1}{2} \left( 
\frac{\partial_i \partial_j }{\partial^2_- } \gamma^{ij}
\right) K_{--} \right) ~ .
\label{gs}
\end{equation}
The field $\gamma^{ij}$ is a traceless symmetric `metric'
with $i = 1, \cdots , D-2$, and
\begin{equation}
K_{\mu \nu} =  \partial_{\mu} \gamma_{ij} 
  \partial_{\nu} \gamma^{ij} ~ ,
\end{equation}
where $\mu = +, - , 1, \cdots , D-2$.  Since our
dilaton-axion system is a nonlinear sigma model
with $SL(2,R)$ invariance, the model (\ref{gs}) 
when $D=4$ should be similar to our system.  This becomes
clear when we consider the following two points.  
First, we introduce two fields 
$C$ and $\bar{C}$ with opposite helicities
for two physical modes of four-dimensional 
gravitons.  The first term of (\ref{gs}) then produces 
$- \partial_+ C \partial_- \bar{C} -
   \partial_+ \bar{C} \partial_- C$ and nothing
else, up to cubic terms, analogous to our action
(\ref{rsugac}).  While this term alone is not 
$J^{i+}$ invariant, adding three extra terms in (\ref{gs})
makes it invariant.  Secondly, the
action of $J^{i+}$ on fields in flat space-time 
is given entirely
by orbital parts with no extra spin contributions; the index
structure of $\gamma_{ij}$ is irrelevant as far as
the action of $J^{i+}$ is concerned, making it easier
to apply to the dilaton-axion system.  The extra term
$H_{GS}$ can be read off from (\ref{gs}) and it can be
shown that $H_{GS}$ makes the action (\ref{rsugac}) Lorentz invariant when
$\mu = 0$.   

The cubic Hamiltonian $H_3$ is consistent
with the light-cone superstring field theory construction
in the pp-wave background of Ref.~\cite{spvo}; 
as a functional of classical fields,  
$H_3$ of supergravity light-cone Hamiltonian
is identical to the flat space ($\mu =0$) result. 
Of course, this does not mean that the matrix elements of 
the quantum operator $H_3$ are $\m$-independent. 
As we have emphasized in the introduction, 
quantum mechanics of free particles ($\m=0$) and 
that of bounded particles ($\m \neq 0$) are quite different. 
For nonzero $\m$, the Hilbert space consists of harmonic oscillator modes 
whereas the $\m = 0$ Hilbert space is a collection of free particles.
As we will see in the next section, the matrix elements of $H_3$ actually 
have some explicit $\m$ dependence due to their dependence 
on the frequency of harmonic oscillators.

To further compare with the analysis of \cite{spvo},
we write down the classical Hamiltonian in momentum space, 
\be
H_3 = 
\int \left( \prod_{r=1}^{3} \frac{d\a_r}{2 \pi} \right) 
\d(\S \a_r)  h_3(\a_r) \t_1 \t_2 \bar{\t}_3 + {\rm c.c.} ~ .
\ee
There are two contributions to $h_3$:
\begin{equation}
h_3^{(0)} =  \frac{\sqrt{2} \kappa}{i}  \frac{1}{4 \alpha_1 \alpha_2} 
( \alpha_1 p_2   - \alpha_2 p_1 )^2
\end{equation}
from $H_3$ other than $H_{GS}$ and
\begin{equation}
h_3^{(GS)} =  \frac{\sqrt{2} \kappa}{i} \cdot \frac{1}{8} 
\left( \frac{1}{\alpha_1^{ 2}} +
\frac{1}{\alpha_2^{ 2} } \right) ( \alpha_1 p_2 - \alpha_2 p_1 )^2 ~
\end{equation}
from $H_{GS}$, which sum up to yield
\begin{equation}
h_3 =  \frac{\sqrt{2} \kappa}{i} \cdot \frac{1}{8} 
   \frac{\alpha_3^{ 2}}{\alpha_1^{ 2} \alpha_2^{ 2} }
  ( \alpha_1 p_2  - \alpha_2 p_1 )^2 .
\label{sugrares}
\end{equation}
Here, $p_t$ are the transverse momenta of the $r$-th particle.

\section{Supergravity sector of light-cone string field theory in a
pp-wave background}

Spradlin and Volovich \cite{spvo} adopted the light-cone string
field theory of \cite{brink} to study string interactions in the
pp-wave background. They showed that the prefactor of
the cubic Hamiltonian $H_3$
is the same as the one in the flat space up to an overall function
$f(\a_1, \a_2, \a_3)$.  We compute the dilaton-axion sector of $H_3$
and compare it to our supergravity result in the previous 
section. 
It will be shown that the function
$f$ is indeed a constant, and the precise value of the
normalization constant will also be determined. We then proceed to 
compute the matrix elements of $H_3$ for three chiral primary 
fields.

\subsection{Dilaton-axion system and normalization of $H_3$}

We begin with a brief review of IIB supergravity in the light-cone
gauge
\cite{extgra} following the notation of \cite{brink}.
A single superfield $\Phi$ contains the 256 physical 
degrees of freedom
of IIB supergravity in the light-cone gauge:
\bea
\Phi(\a, p, \l) &=& \frac{1}{4} \a^2 A(\a, p)
+\frac{4}{8!\a^2} A^*(\a, p) \e^{a \cdots h} \l_a \cdots \l_h
+\frac{1}{4!} A^{abcd}(\a,p) \l_a\l_b\l_c\l_d
\nonumber \\
&& + \frac{1}{4}\a A^{ab}(\a,p) \l_a\l_b
-\frac{1}{6! \a} A^{ab *}(\a,p) \e^{a\cdots h} \l_c \cdots \l_h +
({\rm fermions}) ~ ,
\eea
where $\lambda_a$ is an $SO(8)$ spinor with positive
chirality.  
Only the fields on the first line will concern us in this paper.
The quadratic Hamiltonian in the pp-wave background is given by \cite{metstey}
\bea
H_2 &=& \half \int \frac{d\a}{2 \pi} \frac{d^8 p}{(2 \pi )^8}  d^8\l
\Phi(-\a, -p, -\l) \left( \triangle_B + \triangle_F \right) 
\Phi(\a, p, \l ) \;~ , \\
\triangle_B &=& p^2 - \frac{1}{4} \a^2 \m^2 \frac{\p^2}{\p p^2} , \;\;\;\;\;
\triangle_F = \a \m \l \Pi \frac{\p }{\p \l} ~ , 
\eea
where $\Pi = \gamma^1 \gamma^2 \gamma^3 \gamma^4$ is the 
$SO(4)$ chirality operator. 
Expansion of $H_2$ in component fields is straightforward. The
$SO(8)$ scalars $A, A^*$ give
\be
H_2 = \int \frac{d\a}{2\pi} A^*(-\a) \triangle_B A(\a) ~ .
\ee
This is exactly the same as the quadratic Hamiltonian of $\t$
field in (\ref{sugh2}) upon identifying $\t(\a) = A(\a)$. For the
reasons explained in the previous section and in Ref.~\cite{spvo}, we
expect that `classically' $H_3$ is essentially the same as in the flat
spacetime. In terms of the light-cone superfields, $H_3$ is given
by
\bea
\label{sftcub}
H_3 &=& \CN \sqrt{2} \k \int d\m_3 \; v^{IJ}(\L) \IP^I \IP^J
\Phi(1) \Phi(2) \Phi(3) ,
\\
\label{measure}
d\m_3 &=& \left( \prod_{r=1}^3 \frac{d\a_r}{2 \pi}
 \frac{ d^8p_r}{(2 \pi )^8 } d^8\l_r \right)
\d(\S \a_r) \d^8( \S p_r) \d^8 (\S \l_r), \\
\IP^I &=& \a_1 p_2^I - \a_2 p_1^I, \\
\L^a &=& \a_1 \l_2^a - \a_2 \l_1^a ,
\eea
where we have introduced a normalization constant $\CN$.  
Furthermore,
the prefactor $v^{IJ}$ is given by
\begin{eqnarray}
v^{IJ} & = & \delta^{IJ}
 + \frac{1}{6 \alpha^2} \gamma^{IK}_{ab} \gamma^{JK}_{cd}
\Lambda^a \Lambda^b \Lambda^c \Lambda^d
+ \frac{16}{8! \alpha^4} \delta^{IJ}
\epsilon_{a \cdots h} \Lambda^a \cdots \Lambda^h 
 \nonumber \\
& & 
 - \frac{i}{\alpha} \gamma_{ab}^{IJ}
 \Lambda^a \Lambda^b 
- \frac{4i}{6! \alpha^3} \gamma^{IJ}_{ab} \epsilon_{abc 
\cdots h} \Lambda^c \cdots \Lambda^{h} ~ ,
\label{vij}
\end{eqnarray}
where $\alpha = \alpha_1 \alpha_2 \alpha_3$. 
The terms on the second line are irrelevant for supergravity 
but become important for string states. 

The dilaton-axion part of the interaction Hamiltonian is
obtained from $H_3$ in (\ref{sftcub}) by performing the fermionic 
integrals and collecting terms involving $A$ and $A^*$.  
The $\L^0$ and the $\L^8$ terms of (\ref{vij}) give nonvanishing
answer (the $\L^8$ term producing the result shown below
and the $\L^0$ term producing its complex conjugate): 
\be
H_3 = 12 \CN \sqrt{2} \k
\int d\bar{\m}_3 \left( \frac{\a_3^2}{\a_1^2\a_2^2} \IP^2 \t_1 \t_2
\bar{\t}_3 \right) + {\rm c.c.} ~ ,
\label{whew}
\ee
where $d\bar{\m}_3$ is the bosonic part of $d\m_3$ defined in (\ref{measure}).
We have $H_3$ from (\ref{sugrares}), which was obtained
from the construction of light-cone Hamiltonian from the
covariant action.  After the 90-degree rotation of (\ref{sugrares})
on the complex plane $\tau$, under which $\tau \rightarrow i \tau$ 
and $\bar{\tau} \rightarrow -i \bar{\tau}$ ($H_2$ remains invariant), 
we have
\be
H_3 = \frac{\sqrt{2} \k}{8}
\int d\bar{\m}_3 \left( \frac{\a_3^2}{\a_1^2\a_2^2} \IP^2 \t_1 \t_2
\bar{\t}_3 \right) + {\rm c.c.} ~ .
\label{whew1}
\ee
Comparing (\ref{whew}) and (\ref{whew1}), we find that 
the normalization constant should be
\be
\CN = \frac{1}{3\cdot 2^5} .
\ee
In conclusion, the possible function $f(\alpha_1 , \alpha_2 , \alpha_3 )$
undetermined from the symmetry consideration should be
set to a constant.

\subsection{Matrix elements of $H_3$ for chiral primaries}

\subsubsection{Interaction Hamiltonian}

It is convenient to divide the $SO(8)$ spinor $\l^a$ into two
groups depending on their $SO(4)$ chirality. For example, we may
choose a basis of gamma matrices in which $\Pi = \g^1 \g^2 \g^3
\g^4$ is diagonal, so that $\l^a$s ($a = 1,2,3,4$) have positive
chirality and $\l^{\bar{a}}$s ($\bar{a} =5,6,7,8$) have negative
chirality. In such a basis, the chiral primary field is related to
the components of the superfield as
\be
\label{s1234}
s(\a) = \frac{1}{\sqrt{2}} A^{5678}(\a),
\;\;\;\;\;\;
s(-\a) = [s(\a)]^* = \frac{1}{\sqrt{2}} A^{1234}(-\a)
\;\;\;\;\; (\a>0) ~ .
\ee
The factor of $\sqrt{2}$ was introduced to normalize the
quadractic Hamiltonian in the standard way:
\be
H_2 = \int d\bar{\m}_3
 \;  s(-\a) (\triangle_B -4\m|\a|) s(\a) ~ .
\ee
In order to obtain the cubic Hamiltonian $H_3$ for $s(\a)$, one
has to expand (\ref{sftcub}) and collect $s^3$ terms. It is clear
that only the middle component
\be
\frac{1}{6 (\a_1\a_2 \a_3)^2} \g^{IK}_{[ab} \g^{JK}_{cd]} \L^a
\L^b \L^c \L^d
\ee
of $v^{IJ}$ will contribute to $H_3$ for chiral primaries since
other terms simply do not have the right number of $\l$s to
saturate the $\l$ integral in (\ref{sftcub}). Moreover, the
relation (\ref{s1234}) implies that it is sufficient to pick out
the terms with all
$\L^a$s having the same chirality, i.e., $\L_1\L_2\L_3\L_4 \equiv
\L_{1234}$ and
$\L_5\L_6\L_7\L_8 \equiv \L_{5678}$. 
When the spinor indices are restricted to the same $SO(4)$ 
chirality subspace, say $a= 1,2,3,4$ space, it can be shown that
\be
\label{spid}
\g^{iK}_{[ab} \g^{jK}_{cd]} = \d^{ij} \e_{abcd} ~, \;\;\;\;
\g^{i'K}_{[ab} \g^{j'K}_{cd]} = -\d^{i'j'} \e_{abcd} ~, \;\;\;\;
\g^{iK}_{[ab} \g^{j'K}_{cd]} = 0 ~ .
\ee
First, note that the $\g^{IK}_{ab}$ vanishes if $I$ and $J$ do not
belong to the same $SO(4)$.   This fact can be verified by using 
the aforementioned basis of gamma matrices where the matrix
$\Pi$ is diagonal.  Second, since there are only four
components of a spinor with the same chirality, the LHS of
(\ref{spid}) must be proportional to $\e_{abcd}$.
With a suitable choice of the basis for $\g$ matrices, 
one can use the self-dual property of $\g^{JK}_{ab}$
\be
\g^{ik}_{ab} = \half \e_{abcd} \g^{ik}_{cd}, \;\;\;\;\;
\g^{i'k'}_{ab} = - \half \e_{abcd} \g^{i'k'}_{cd} 
\ee
to show that the RHS of (\ref{spid}) is correctly normalized 
and that there is a relative minus sign between the two $SO(4)$s. 

Without loss of generality, one may assume that $\a^{(1)} ,
\a^{(2)} > 0$ and $\a^{(3)} < 0$.
In such a case, one encounters the following expression,
\be
\label{llll}
\L_{1234} (\l^{(1)} + \l^{(2)})_{1234},
\ee
where $\L = \a_1 \l^{(2)} - \a_2 \l^{(1)}$. Only the terms
proportional to $\l^{(1)}_{1234} \l^{(2)}_{1234}$ saturate the
$\l$-integral. Collecting all such terms, one finds
\be
(\a_1 + \a_2)^4 \l^{(1)}_{1234} \l^{(2)}_{1234}
= \frac{1}{16} (|\a_1| + |\a_2| + |\a_3|)^4
 \l^{(1)}_{1234} \l^{(2)}_{1234} ~ .
\ee
All in all, we find that $H_3$ for chiral primaries is given by
\be
\label{cp3}
H_3 = \frac{2^{3/2} \sqrt{2}\k}{3\cdot 2^7} 
\int d\bar{\m}_3 \left( 
\frac{(|\a_1| + |\a_2| + |\a_3|)^4}{\a_1^2 \a_2^2 \a_3^2}
(\IP_{\parallel}^2  - \IP_{\perp}^2 ) s_1 s_2 s_3 \right) ~ .
\ee
where $\parallel$ and $\perp$ denote 
the four transverse directions coming from AdS$_5$ and 
the other four from $S^5$, respectively.

\subsubsection{Quantization}

Consider a massless scalar in the pp-wave background 
in the light-cone gauge 
\footnote{
For simplicity, we consider a massless scalar in place of the 
chiral primary field $s$. The difference between the two fields 
will not be important in what follows.
}
\bea
S &=& \int d^{10}x \sqrt{-g} \left( -\half (\del\phi)^2 \right)
\nonumber \\
&=& \int dx^+ dx^- d^8\vx 
\left( \p_+ \phi \p_- \phi -\frac{1}{4} \m^2 x^2 (\p_- \phi)^2 -
(\p_I \phi)^2 \right)~ .
\eea
The Dirac quantization procedure for a constrained system gives
the quantum commutation relation
\be
\left[ \phi(x^-, \vx), \p_- \phi(y^-, \vy) \right]
= \frac{i}{2} \d (x^- - y^-) \d^8 (\vx - \vy) ~ .
\ee
The normalizable on-shell wavefunctions are labeled by $\a, \vn$
\be
\phi(x^+, x^-, \vx) = e^{-iEx^+ - i \a x^-} f_{\vn} (\vx) ~ .
\ee
Here, $f_{\vn}$ is the $\vn$-excited state wave function of an eight
dimensional harmonic oscillator with $\w = \m|\a|/2$ and the
energy is given by
$E=\m (||\vec{n}|| +4)$ where $||\vec{n}|| \equiv \sum_{I=1}^8 n_I$.
We find it convenient to focus on the $x^-$ (or $\a$ in the
momentum space) dependence, suppressing the transverse directions
as long as no confusion arises. Going to the momentum space only in the
$x^-$ direction via Fourier-transform
\be
\phi(x^-) = \int \frac{d\a}{2\pi} \phi(\a) e^{-i\a x^- },
\ee
the commutation relation becomes
\be
\left[ \phi(\a_1) , \phi(\a_2) \right] = \frac{1}{2\a_1} 2\pi
\d(\a_1 + \a_2) ,
\ee
which is solved by
\be
\phi(\a) = \frac{\sqrt{2\pi}}{\sqrt{2} \a} \sum_{\vn} a_{\vn}(\a)
f_{\vn}(\vx) \;\;\;\;\;\;\;\;\;\; (\a>0),
\ee
where the oscillators satisfy
\be
\left[ a_{\vec{m}}(\a_1), a^\dagger_{\vec{n}}(\a_2) \right]
= \d_{\vec{m}, \vec{n}} \a_1 \d (\a_1 - \a_2) .
\ee
The corresponding single particle states are normalized
accordingly,
\be
\langle \a_1, \vec{m} | \a_2, \vec{n} \rangle  =
\d_{\vec{m}, \vec{n}} \a_1 \d (\a_1 - \a_2).
\ee
In terms of the oscillators the quadratic Hamiltonian is expressed
as
\be
H_2 = \m \sum_{\vec{n}} \int \frac{d\a}{2\pi \a}
a^\dagger_{\vec{n}}(\a) a_{\vec{n}} (\a) ( ||\vec{n}|| +4 ).
\ee

\subsubsection{Matrix elements of $H_3$}

Suppose we have a cubic Hamiltonian of the form
\be
H_3 = \int \frac{d\a_1 d\a_2 d\a_3}{(2\pi)^3} (2\pi) \d (\S \a_r)
h_3 (\a_r) s(\a_1) s(\a_2) s(\a_3) ~ .
\ee
It is easy to show that the matrix element is given by
\be
\label{h3el}
\langle 3| H_3 |1,2 \rangle = \frac{2\pi \cdot 3! }{2^{3/2} (2\pi)^{3/2}}
h_3(\a_r) \d(\S \a_r ) \times (\vn_r| E(b_r) |0) ~ .
\ee
Note that we have included the symmetry factor $3!$ and factors of
$(2\pi)$s and $\sqrt{2}$s that we introduced when writing
fields in terms of creation/annihilation operators. The
dimensionless factor $(\vn_r| E(b_r) |0)$ arises from the overlap
integral of the wave function in the transverse directions. In the
language of quantum mechanics of a harmonic oscillator, we have the
following expressions:
\bea
f_{\vn} (\vx) &=& (\vn | \vx) = (\vn | \hat{O}(\vx) |0), \\
\hat{O}(\vx) &=& \left(\frac{\m \a}{2\pi}\right)^2
\exp\left( -\half \va^\dagger \cdot \va^\dagger +
\sqrt{\m |\a|} \vx\cdot\va^\dagger - \frac{1}{4} \m |\a| \vx^2 \right).
\eea
Integration over the eight transverse directions becomes a simple
Gaussian integral which produces
\be
\frac{2^4}{(2\pi)^2}
\frac{\m^2 (\a_1\a_2\a_3)^2}{(|\a_1|+|\a_2|+|\a_3|)^4}
\times (\vn_r| E(b_r) |0),
\ee
where the operator $E(b_r)$ defined by 
\be
E \equiv \exp \left( 
\half \sum_{r,s=1}^{3} a^\dagger_{(r)} M^{rs} a^\dagger_{(s)} 
\right), \;\;\;\;\; 
M^{rs} \equiv \left(\matrix{ 
1 - b_1^2 & - b_1 b_2  & -b_1 b_3 \cr
- b_1 b_2 & 1 - b_2^2 & - b_2 b_3 \cr
- b_1 b_3 & - b_2 b_3 & 1 - b_3^2 }
\right) , 
\ee
and $b_r \equiv \sqrt{2|\a_r|/(|\a_1|+|\a_2|+|\a_3|)}$
is precisely the same as the operator
$E_a^0$ defined in Eqs.~(4.10-11) of \cite{spvo}.
Before using $h_3$ for the chiral primaries in (\ref{cp3}), 
it is useful to realize that
within the momentum integral one can write
\be
\IP_{\parallel}^2 - \IP_{\perp}^2 = 
\a_1\a_2\a_3 (E_{123}^{\parallel} - E_{123}^{\perp}) ,
\ee
where 
$E_{123}^{\parallel} \equiv E_{3}^{\parallel} - 
E_{1}^{\parallel} - E_{2}^{\parallel}$
and
$E_{123}^{\perp} \equiv E_{3}^{\perp} - 
E_{1}^{\perp} - E_{2}^{\perp}$
are the contributions to the energy difference 
from the two $SO(4)$ directions.
This identity \cite{spvo} follows from the definition of 
the free part of the light-cone Hamiltonian for a chiral primary field,
\be
\label{hhh}
H = \frac{1}{\a}\left( p^2 + \frac{1}{4}(\m\a)^2 x^2 \right) - 4\m ,
\ee
and the fact that the sum of three $x$ terms vanishes 
due to momentum conservation ($\a_1 + \a_2 + \a_3 = 0$). 
The constant term in (\ref{hhh}) originates from 
the $\triangle_F$ term in (\ref{sugh2}) and 
cancels the zero point energy of the bosonic harmonic oscillator part.
Since one has only the difference between the two $SO(4)$ directions, 
the zero point energy cancels out automatically and the term $4\mu$ 
is irrelevant. Note also that in (\ref{h3el}), 
$\a_1$ and $\a_2$ are positive while $\a_3 = -(\a_1+\a_2)$ is negative. 
Since the eigenvalue of Hamiltonian (\ref{hhh}) has the same sign as $\a$, 
if one defines $E_r$ to be the absolute value of the energy of the $r$-th 
state, one gets $ \IP_{\parallel}^2 = \a_1\a_2\a_3 E_{123}^{\parallel}$ 
and  $ \IP_{\perp}^2 = \a_1\a_2\a_3 E_{123}^{\perp}$ 
Taking everything into account, we insert the $h_3$ of (\ref{cp3}) 
into (\ref{h3el}) to find  
\be
\label{shinnanda}
\langle 3| H_3 |1,2 \rangle = \frac{\pi}{2}
g\a'^2 \m^2 \a_1 \a_2 \a_3
(E_{123}^{\perp} - E_{123}^{\parallel}) \d(\S \a_r )
\times (\vn_r| E(b_r) |0) ~ .
\ee
The authors of Ref.~\cite{harvmit} 
made a proposal for the matrix elements 
that is supposed to be valid when 
the energy difference between the in-state 
and the out-state is nearly zero. 
\footnote{
For the supergravity modes, the energy difference is 
always an integer (times $\m$), so one may think that 
we are comparing a zero with another zero in the rest of the section. 
A related subtlety that arises in the computation of extremal correlators 
in AdS \cite{3pt} was circumvented by using analytic continuation. 
In the same spirit, we use analytic continuation to 
give a meaning to the coefficient multiplying the `zero.'
}
In the Yang-Mills variables, the proposal reads
\be
\langle 3 | H_3 | 12 \rangle = \m (\D_3 - \D_2 - \D_1) 
C_{123} \d_{J_1+J_2, J_3}.
\ee
The value of $C_{123}$ depends on the states 
$\{ |1\rangle, |2\rangle, |3\rangle \}$. 
To compare with the results of Ref. \cite{harvmit}, 
we introduce the following operators in Yang-Mills 
and their couterparts in supergravity, 
\bea
A^J = \frac{1}{\sqrt{J N^J}} \Tr Z^J \;\;\;\; 
&\leftrightarrow& \;\;\;\; | \a, \vec{0} \rangle , 
\\
B^J =
\frac{1}{\sqrt{N^{J+1}}} \Tr ( \phi Z^{J}) \;\;\;\;
&\leftrightarrow& \;\;\;\; a_\phi^\dagger | \a, \vec{0} \rangle  ,
\\
C^J = \frac{1}{\sqrt{J N^{J+2}}} \sum_{l=0}^{J}
\Tr( \phi Z^l \psi Z^{J-l} )  \;\;\;\;
&\leftrightarrow& \;\;\;\; 
a_\psi^\dagger a_\phi^\dagger | \a, \vec{0} \rangle .
~ ,
\eea
There are four processes with $E_{123} \approx 0$. 
The value of $C_{123}$ for each case can be read off from section 3.2 of 
Ref. \cite{harvmit}:
\bea
AA \goto A \;\; &:& \;\; C_{123} = \frac{\sqrt{J_1J_2J_3}}{N} 
\equiv C_{123}^{(0)}  ,
\\
AB \goto B \;\; &:& \;\; C_{123} = C_{123}^{(0)} 
\times \sqrt{\frac{J_2}{J_3}} , 
\\
AC \goto C \;\; &:& \;\; C_{123} = C_{123}^{(0)} \times \frac{J_2}{J_3} ,
\\
B_{\phi}B_{\psi} \goto C \;\; &:& \;\; C_{123} = C_{123}^{(0)} 
\times \frac{\sqrt{J_1J_2}}{J_3} .
\eea
Note that since $J$ is proportional to $\a$, 
the relative factor matches precisely with $(\vn_r | E(b_r) | 0)$. 
Therefore, it suffices to check the corrspondence for 
the case $AA \goto A$. 

Following Ref. \cite{harvmit}, 
we switch from the unit normalization $\langle i| j \rangle = \d_{ij}$ 
to a normalization suitable for the continuum limit, 
$\langle i| j \rangle = J_i \d_{ij} = \a_i \d(\a_i-\a_j)$, 
and use $J= \m R^2 \a /2$ and $R^4 = 4\pi g_s \a'^2 N$ to find  
\be
\label{ymsg}
\langle 3| H_3 |1,2 \rangle = \pi
g\a'^2 \m^2 \a_1 \a_2 \a_3
E_{123} \d(\S \a_r )
\ee
which indeed agrees with the supergravity result (\ref{shinnanda}) 
with $\vn_r = 0$ (up to a numerical factor of two).
In fact, the authors of Ref. \cite{harvmit} 
considered only the processes with strictly vanishing $E_{123}^{\parallel}$ 
and slightly nonzero $E_{123}^{\perp}$. Our result (\ref{shinnanda}) 
suggests that the Yang-Mills computation should also 
distinguish the two $SO(4)$ directions and 
$E_{123}$ in (\ref{ymsg}) be replaced by 
$E_{123}^{\perp} - E_{123}^{\parallel}$.
\footnote{
After submitting this paper, 
we were informed that the relative minus sign between the
two $SO(4)$ directions and its implications 
have been noticed independently in Refs. \cite{harvmit,lmp}.
}
It would be interesting to verify this expectation explicitly
on the Yang-Mills side.

\section*{Acknowledgement}

We are grateful to Jin-Ho Cho, Shiraz Minwalla, Jongwon Park and
Hyeonjoon Shin for useful discussions, 
and Peter Lee for pointing out an error in the orignial version 
of this paper.
We also thank the organizers of the KIAS Workshop on Strings and Branes
for hospitality and for an opportunity to present this work.
Y. Kiem is especially grateful to
Seungjoon Hyun for pointing out a subtlety in 
Section 2 regarding the zero mode treatment.  
The work of Y. Kiem and Y. Kim was supported by 
Korea Research Foundation Grant KRF-2001-015-DP0082.  
The work of S. Lee was supported by the CNNC.
The work of J. Park was supported by the 
POSTECH BSRI research fund 1RB0210601.


%

\begin{appendix}

\section{AdS computation}

In this appendix, we compute the matrix elements 
for three chiral primary states $(\D = J)$. 
The AdS supergravity
action for the fields corresponding to the chiral primaries is
known to be \cite{3pt}
\be
\label{cubact}
S = \int d^5 x \sqrt{-g} \left(
 - \del s^I \del \bar{s}^I  - m_I^2  |s^I|^2 
 -\frac{1}{2} G_{IJK} (s^I s^J \bar{s}^K  + {\rm c.c.} ) 
 \right) ~ .
\ee
When all three $s$ fields have $\D - J = 0$ , 
the coupling constant is
\bea
G_{123} &=& (\sqrt{2} \k) 
\frac{2^{1/2} \sqrt{J_1J_2J_3 (J_3^2-1)(J_3+2)}(J_3-2) \D_{123}}
{\sqrt{(J_1^2-1)(J_2^2-1)(J_1+2)(J_2+2)}}
\times F_{123} ~ ,  
\\
F_{123} &=& \frac{1}{\sqrt{2\pi^3}} 
\frac{\sqrt{(J_1+1)(J_1+2)(J_2+1)(J_2+2)}}
{\sqrt{(J_3+1)(J_3+2)}} ~ .
\eea
The coefficient $F_{IJK}$ comes from the overlap integral 
of spherical harmonics on $S^5$. This factor gets modified 
when one considers fields with different values of 
$\D - J$, but as we discussed in section 4, 
the change again matches $(\vn_r|E(b_r)|0)$. 

It is straightforward to compute the matrix element of the cubic
Hamiltonian from the action (\ref{cubact}). Each on-shell
wavefunction to the linearized equation of motion corresponds to a
single particle state in the quantum theory. For simplicity, we
restrict ourselves to the ground state ($E = \D$) for which the
normalized wave-packet is given by
\be
s = \frac{\sqrt{\D(\D-1)}}{\pi \; (\cosh\rho)^\D } ~ ,
\ee
where $\rho$ is the radial variable of the AdS global coordinates.
Following the standard recipe of quantum field theory, we find
\bea
\langle 3| H_3 | 12 \rangle
= \frac{1}{2^{3/2} \pi}
\frac{\sqrt{(\D_1-1)(\D_2-1)(\D_3-1) } }
{ (\D_3 -1) (\D_3-2) } \times G_{123} ~ .
\eea
Using the $G_{123}$ written above, 
we find the following remarkably simple result:
\be
\langle 3| H_3 | 12 \rangle =
(\D_3 - \D_1 - \D_2) \frac{\sqrt{\D_1 \D_2 \D_3}}{N} ~ ,
\ee
which is in agreement with the proposal of \cite{harvmit}.

All of the above appear sensible, but we have to admit that 
there are a few reasons to doubt the validity of this computation. 
First, in the process of obtaining the cubic term in the action 
(\ref{cubact}), one made a nonlinear field redefinition 
in which the on-shell condition $\del^2 s = m^2 s$ was used. 
In particular, on-shell condition removed cubic 
couplings with time derivatives, which may cause trouble in quantization.
Second, the method used in this appendix gives 
answers that are finite in the pp-wave limit only when 
$\D_{123} = 0$. If the cubic Hamiltonian we used were
the correct one, the matching between the AdS Hamiltonian and 
the pp-wave Hamiltonian would be valid for arbitrary values of 
$\D_{123}$.

\end{appendix}

\newpage

\rnc{\baselinestretch}{1.20}

\end{document}